\documentstyle[11pt,fleqn]{article}
\input{psfig.tex}
\title{Comments on "Sudden future singularities "} 
\author{A. Kwang-Hua CHU \thanks{Present Address :
Department of Physics, Xinjiang University, 14, Shengli Road,
Urumqi 830046, PR China.  }}  
\date{P.O. Box 30-15, Shanghai 200030, PR China}
\begin{document}           
\maketitle
\begin{abstract}
We make comments on the letter by Barrow [{\it Class. Quantum
Grav. } {\bf 21} (2004) L79] since we cannot observe a singularity
for $p(t)$ as $t\rightarrow t_s$. Instead, there might be a
singularity for $\rho(t)$ for $t\sim 0$ case. \newline

\end{abstract}
\doublerulesep=6mm    %
\baselineskip=6mm
\bibliographystyle{plain}               
\noindent

Barrow just presented an interesting letter which shows that a
singularity can occur at a finite future time in an expanding
Friedmann universe even when $\rho > 0$ and $\rho + 3\,p > 0$ [1].
He has identified conditions that are sufficient to avoid the
appearance of this type of sp-curvature singularity at finite
future time. These results can either be used as a more
fine-grained method to search for finitetime singularities and
evaluate their likelihood of occurrence with realistic fluids or,
if such singularities are regarded as causal pathologies, to
impose realistic conditions which exclude them by fiat. \newline
Consider the Friedmann universe with expansion scale factor
$a(t)$, curvature parameter $k$, and Hubble expansion rate $H =
\dot{a}/a$; then ($8\pi G = c = 1$) the Einstein equations reduce
to
\begin{equation}
 3 H^2=\rho-\frac{k}{a^2}, \hspace*{24mm} \dot{\rho}+3 H
 (\rho+p)=0,
\end{equation}
and
\begin{equation}
 \frac{\ddot{a}}{a}=-\frac{\rho+3 p}{6}.
\end{equation}
He first looked informally at whether it is possible for any type
of singularity to develop in which a physical scalar becomes
infinite at a finite future comoving proper time $t_s$ , when,
say, $a(t) \rightarrow a(t_s) \not = 0$ or $\infty$, and $H(t)
\rightarrow H_s < \infty$ and $H_s > 0$. \newline Guided by these
heuristic arguments, he then constructed an explicit example by
seeking, over the time interval $0 < t < t_s$ , a solution for the
scale factor $a(t)$ of the form $a(t) = A + B \,t^q + C(t_s -
t)^n$, where $A > 0$,$B > 0$, $q > 0$, $C$ and $n > 0$ are free
constants to be determined. \newline He can obtain (Eq. (14) in
[1])
\begin{displaymath}
 \frac{\ddot{a}}{a} \rightarrow -\infty,
\end{displaymath}
whenever $1 < n < 2$ and $0 < q \le 1$; the solution exists on the
interval $0 < t < t_s$ . Hence, as $t \rightarrow t_s$ we have $a
\rightarrow a_s$ ; $H_s$ and $\rho_s > 0$ (for $3 q^2(a_s - 1)^2
t^{-2}_s > -k$) are finite but $p_s \rightarrow \infty$. When $2 <
n < 3$, he noted that $\ddot{a}$ remains finite but $d^3 a/dt^3
\rightarrow \infty$ as $t \rightarrow t_s$ as $p_s$ remains finite
but $\dot{p}_s \rightarrow \infty$. By contrast, there is an
initial all-encompassing strong-curvature singularity, with $\rho
\rightarrow ¡ú\infty$ and $p \rightarrow \infty$, as $t
\rightarrow 0$. However, from Eqs. (14) and (4) of [1], he can see
that $\rho$ and $\rho +3p$ remain positive. \newline After
considering two approximate forms of $a(t)$ as $t\rightarrow 0$
and  $t_s$, he then claimed that this specific family of solutions
shows that it is possible for an expanding universe to develop a
'big-rip' singularity at a finite future time even if the matter
fields in the universe satisfy the strong-energy conditions $\rho
> 0$ and $\rho + 3 p > 0$. \newline
The present author would like to make comments about his
derivations (especially about the equation (4) and its subsequent
variations in [1]) and his results, say, equations (11) or (13) in
[1]. Our second remark could be induced by the first remark!
\newline The first remark has been partially addressed, as Barrow
noticed that, there are two independent Friedmann equations for
the three quantities $a$, $p$ and $\rho$. In the absence of a
relation between any two of them there is an unconstrained degree
of freedom. If we introduce an equation of state $p(\rho)$ which
bounds the pressure by some well-behaved function of the density¡ª
for example, $p < C \rho$ with $C > 0$-then the pressure
singularity at a finite future time is eliminated if we require
$\rho > 0$ and $\rho +3p> 0$. An upper bound on the ratio $p/\rho$
corresponds to a bound on signal propagation of waves associated
with small changes in pressure.
\newline The second remark will be demonstrated by my illustration
of figures once we plot $\ddot{a(t)}/a(t) [\propto p(t)$] and
$\rho(t)$ with respect to $t$ (from the equation (13) of [1]). As
Barrow mentioned, $1 < n < 2$ and $0 < q \le 1$, thus we choose
$n=1.5$, $q=0.8$, and $t_s =10$. But, the trouble is that $a_s$ is
unknown. We then try three cases : $a_s =1.00,1.05,50.00$. We can
easily observe from Fig. 1 that there might be a singularity for
$\rho(t)$ but not for $p(t)$ as $t \sim 0$ once $a_s \sim 1$.  We
cannot observe a singularity for $\rho(t)$ or $p(t)$ once
$t\rightarrow t_s$ which Barrow claimed in [1] that there will be
a singularity for $p(t)$! These can be demonstrated in Figs. 1, 2
and 3 for either $t_s =10$ or $t_s=100$ once $t\rightarrow t_s$.
\newline
Let us consider the expression of $\ddot{a}(t)/a(t)$, the important term is
$t_s -t = K\, t_s$. If $K \ll 1$, then there will be no singularity
for $p(t)$ as $t \rightarrow t_s$ once $t_s^2 \times K^{2-n} \sim O(1)$!
\newline
To conclude in brief, our remarks as demonstrated in Figs. 1, 2
and 3 could not only be valid to the false statement which Barrow
made in [1] but also be possible be valid to his subsequent paper
(quite similar works) in [2].

\newpage

\vspace{5mm}
\psfig{file=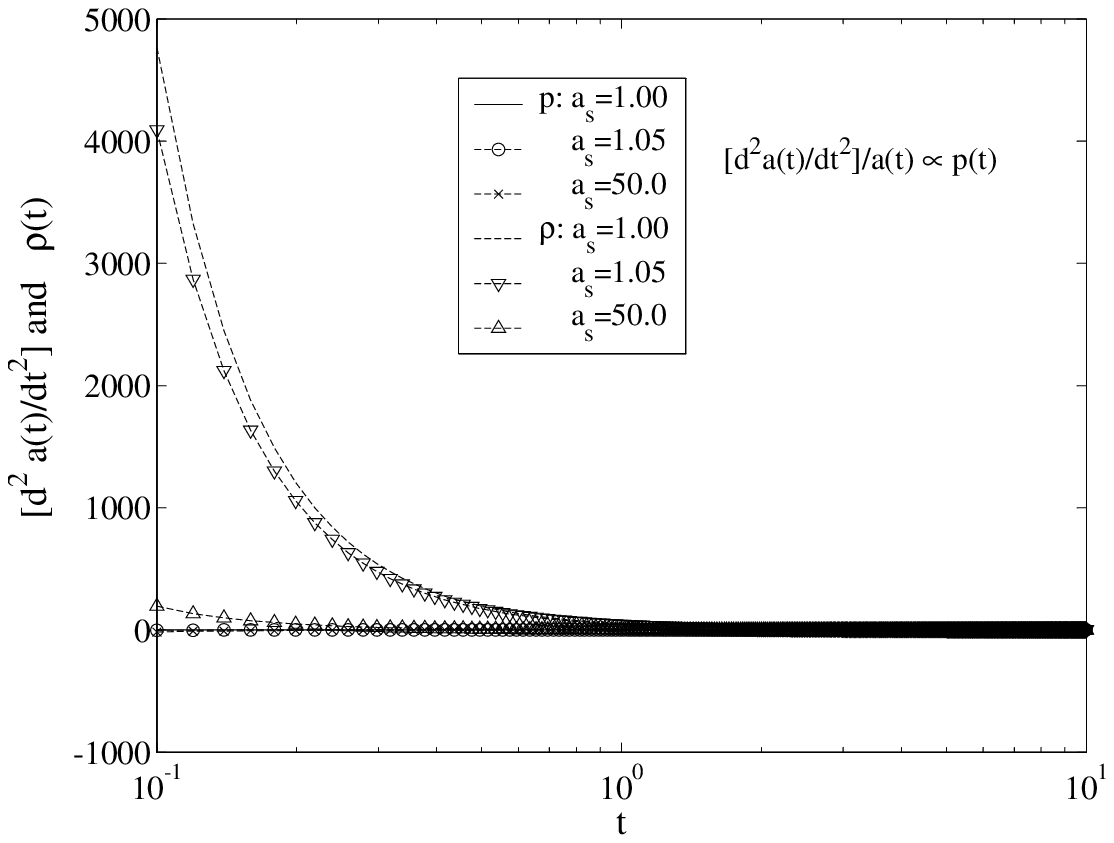,bbllx=0cm,bblly=14cm,bburx=22cm,bbury=24cm,rheight=9cm,rwidth=9cm,clip=}
\begin{figure}[h]
\hspace*{6mm} {\small Fig. 1 \hspace*{2mm} $ p(t) [\propto
\ddot{a}(t)/a(t)] $ and $\rho(t)$ vs. $t$. Here, $n=1.5$, $q=0.8$,
$t_s =10.0$.
\newline \hspace*{6mm} There might be a singularity for $\rho(t)$
but not for $p(t)$ as $t\sim 0$ \newline \hspace*{6mm} (case of
$a_s \sim 1$; but not for $t\rightarrow t_s$).}
\end{figure}

\vspace{15mm}

\psfig{file=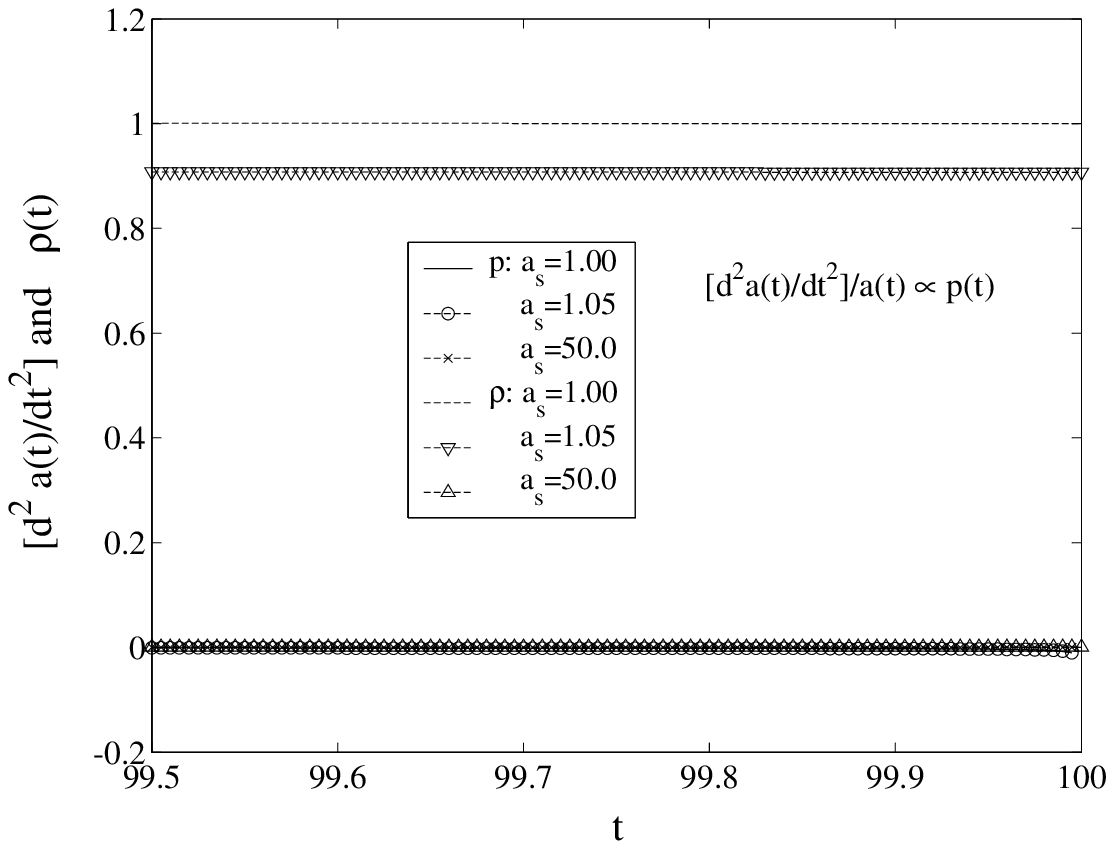,bbllx=0cm,bblly=14.4cm,bburx=20cm,bbury=23.8cm,rheight=8.8cm,rwidth=8.8cm,clip=}
\begin{figure}[h]
\hspace*{6mm} {\small Fig. 2 \hspace*{2mm} $ p(t) [\propto
\ddot{a}(t)/a(t)] $ and $\rho(t)$ vs. $t$. Here, $n=1.5$, $q=0.8$,
$t_s =100$.
\newline \hspace*{6mm} There is no singularity for $p(t)$ or
$\rho(t)$ once $t\rightarrow t_s (\equiv 100)$.}
\end{figure}

\newpage
\psfig{file=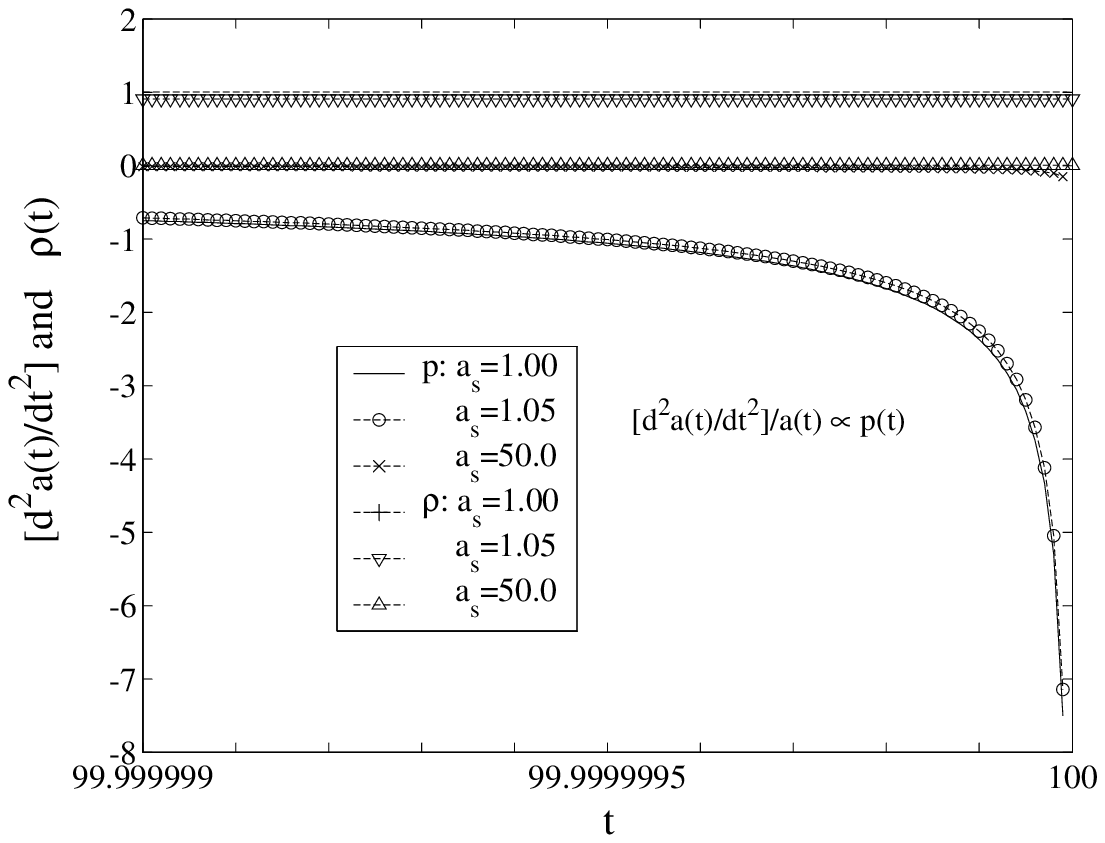,bbllx=0cm,bblly=14.4cm,bburx=20cm,bbury=23.8cm,rheight=8.8cm,rwidth=8.8cm,clip=}
\begin{figure}[h]
\hspace*{6mm} {\small Fig. 3 \hspace*{2mm} $ p(t) [\propto
\ddot{a}(t)/a(t)] $ and $\rho(t)$ vs. $t$. Here, $n=1.5$, $q=0.8$,
$t_s =100$.
\newline \hspace*{6mm} There might be no singularity for $p(t)$ or
$\rho(t)$ once $t\rightarrow t_s (\equiv 100)$.}
\end{figure}
\end{document}